\documentclass[twocolumn,showpacs,preprintnumbers,amsmath,amssymb]{revtex4}
\usepackage{graphicx,subfigure}
\usepackage{color}
\newcommand {\bea} {\begin{eqnarray}}
\newcommand {\eea}{\end{eqnarray}}
\newcommand {\mup}{m_{\uparrow}}
\newcommand {\md}{m_{\downarrow}}
\begin{document}
\draft
\title{Virial expansion with Feynman diagrams}
\author{X. Leyronas}
\affiliation{
Laboratoire de Physique Statistique, Ecole Normale Sup\'erieure, UPMC
Univ Paris 06, Universit\'e Paris Diderot, CNRS, 24 rue Lhomond, 75005 Paris,
France.}

\begin{abstract}
We present a field theoretic method for the  calculation of the second and third virial coefficients $b_2$ and $b_3$ of $2$-species fermions interacting via a contact interaction. The method is mostly analytic. We find a closed expression for $b_3$
in terms of the $2$ and $3$-body $T$-matrices. We recover numerically, at unitarity, and also in the whole BEC-BCS crossover, previous numerical results for the third virial coefficient $b_3$.
\end{abstract}

\maketitle

\section{Introduction}
The topic of interacting ultracold fermions has witnessed some spectacular experimental
developments since the first formation of a molecular Bose Einstein Condensate
\cite{becmol}. The Bose Einstein Condensate (BEC)-Bardeen Cooper Schrieffer (BCS)
crossover has been the subject of numerous experimental and theoretical works \cite{gps}. Recently, the authors of Ref.\cite{eosLi6ens} have found an accurate and elegant way of measuring the equation of state of an ultracold gas.  In particular, they were able to measure the equation of state of a mixture of $^6Li$ atoms in two internal states $\uparrow$ and $\downarrow$. In this experiment, the pressure $P$ was measured as a function of the chemical potential $\mu$ and the temperature $T$. 
The so called {\it unitary limit} (see below), where the interactions are known to be strong, was studied, starting from high temperature down to low temperature (close to the superfluid transition). In the high temperature regime, the data was compared to a {\it virial expansion} \cite{llphysstat}
\bea
P(\mu,T)&=&2\,T(\Lambda_T)^{-3}\left\{z+b_2 z^2+b_3 z^3+\cdots+b_n\,z^n
+\cdots
\right\}\nonumber\\
&&\label{eqviriel}
\eea
where $\Lambda_T=\sqrt{\frac{2\pi}{m\,T}}$ is  the thermal wavelength (we take $\hbar=1$ and the Boltzmann constant $k_B=1$). $z=e^{\beta\mu}$ is the fugacity and 
$\beta=T^{-1}$. In this way, the authors of Ref.\cite{eosLi6ens} were able to extract the numerical values of $b_3$ and $b_4$.

The theoretical calculation of virial coefficients started in the late $30$'s with the work of Beth and Uhlenbeck \cite{bethuhle} on the calculation of $b_2$ (see also \cite{llphysstat} and \cite{prackl}). The problem of computing $b_3$ was considered later in Ref.\cite{paisuhl}.
In the context of ultracold atoms, $b_3$ for ultracold bosons was studied in Ref.\cite{bedaquerupak} and later for ultracold fermions in Refs.\cite{rupak,hld,kaplansun,rdblume}. It is the goal of this paper to present a diagrammatic method to calculate $b_3$. Our results are in full agreement with Refs.\cite{hld,rdblume} and therefore disagree with Ref.\cite{rupak}. The idea of the method is that one needs to include the $3$-body problem in the {\it many}-body problem if one wants to calculate $b_3$. In some sense, this is similar to the zero temperature low density expansion of Refs.\cite{prllhy,prallhypol}. In this work too, we will need to use the $T$-matrix of the $3$-problem. Since the $3$-body problem is basically solved \cite{stm,petrov,pra4par}, we can calculate $b_3$.

The paper is organized as follows. In section \ref{general}, the general formalism is introduced. The technical key point is to work in (imaginary) time, rather than with frequencies. In section \ref{b2}, we apply the method to the determination of $b_2$ and recover in a rather efficient way the Beth-Uhlenbeck result \cite{bethuhle}. Section \ref{b3}
is the main part of the paper, and it explains how to determine diagrammatically $b_3$.
Several technical details are given in the Appendices.
The main analytical results are given in Eqs.(\ref{eqres332a}),(\ref{eqres332b}),(\ref{eqres32a}) and 
(\ref{eqb3fin}). Numerical results are shown in Fig.\ref{figb3} and Eq.(\ref{eqb3lunum}).
Finally, we conclude in section \ref{conclusion}.

\section{General formalism}\label{general}
We describe now the method we develop to get an expansion in power of the fugacity. We first consider the Green's function for free fermions in imaginary time
(for a general reference on diagrammatic techniques, see Ref.\cite{AGD}).
We have 
\bea
G^{0}({\bf p},\tau)&=&e^{-(\varepsilon_{{\bf p}}-\mu)\tau}
\left\{-\Theta(\tau)+n_F(\varepsilon_{{\bf p}}-\mu)
\right\}\label{eqG0in}
\eea
 with $n_F(x)=(e^{\beta\,x}+1)^{-1}$ and 
$\varepsilon_{{\bf p}}={\bf p}^2/(2 m) $ is the kinetic energy of a fermion of mass $m$.
$\Theta(x)$ is the Heaviside function.
In the high temperature limit we consider, the fugacity $z=e^{\beta\mu}$ is smaller than one , and we can expand the Fermi-Dirac distribution according to
\bea
n_F(\varepsilon_{{\bf p}}-\mu)&=&\frac{z\,e^{-\beta\varepsilon({\bf p})}}{1+z\,e^{-\beta\varepsilon({\bf p})}}\nonumber\\
&=&\sum_{n\geq 1}z^{n}(-1)^{n-1}e^{-n\beta\varepsilon({\bf p})}\label{eqnf}
\eea
From Eq.\ref{eqG0in} and \ref{eqnf}, we find an expansion in powers of the fugacity for the free fermions Green's function
\bea
G^{0}({\bf p},\tau)&=&e^{\mu\tau}\left[
\sum_{n\geq 0} 
G^{(0,n)}({\bf p},\tau)\,z^n
\right]\label{eqG0}
\eea
 We have defined
\bea
G^{(0,0)}({\bf p},\tau)&=&-\Theta(\tau)e^{-\varepsilon_{{\bf p}}\tau}\\
G^{(0,n)}({\bf p},\tau)&=&(-1)^{n-1}e^{-\varepsilon_{{\bf p}}\tau}e^{-n\beta\varepsilon_{{\bf p}}},\,n\geq 1\label{eqg0nf}
\eea
Therefore $G^{(0,0)}$ is {\it retarded}, while $G^{(0,n)}$, for $n\geq 1$, is {\it not retarded}.
Notice that $G^{(0,n)}$ does not depend on the chemical potential $\mu$.
Diagrammatically, since $G^{(0,0)}$ is a retarded function, we represent it as a line with an arrow going {\it from left to right} if increasing time goes to the right (this is the Green's function of a particle in vacuum).
On the other hand, $G^{(0,n)}$, for $n\geq 1$, is not retarded, and we represent it as a  $n$-times slashed line, which can be oriented from left to right or {\it vice versa}. 
In order to calculate the virial coefficients, we expand the density $n(\mu,T)$ (per spin species for spin $1/2$ fermions) in power of the fugacity $z=e^{\beta\mu}$. 
\bea
n(\mu,T)&=&(\Lambda_T)^{-3}\left\{z+2 b_2 z^2+3 b_3 z^3+\cdots+n b_n z^n+\cdots
\right\}\nonumber\\
&&
\eea
which comes from Eq.(\ref{eqviriel}) and the Gibbs-Duhem identity $n=\left(\partial P/\partial \mu\right)_{T}/2$.
The density per spin is related to the full Green's function $G({\bf k},\tau)$ via the equation
\bea
n(\mu,T)&=&\sum_{{\bf k}} G({\bf k},\tau=0_{-})
\eea 
where we denote $\sum_{{\bf k}}=\int\frac{d^{3}{\bf k}}{(2\pi)^3}$ in $3$ dimensions.

It is convenient to define the densities $\delta n^{(p)}$ such that
\bea
n(\mu,T)&=&\delta n^{(1)}z+\delta n^{(2)}z^2+\delta n^{(3)}z^3+\cdots
\eea

The principle of the method is the following : in order to calculate $\delta n^{(p)}$, we need to find all the diagrams for $G_{\uparrow}({\bf k},0_{-})$ of order $z^p$. Those are the diagrams with one $G^{(0,p)}$, or one $G^{(0,p-1)}$ and one $G^{(0,1)}$, 
or one $G^{(0,p-2)}$ and one $G^{(0,2)}$, or one $G^{(0,p-2)}$  and two $G^{(0,1)}$'s
etc...
Of course, the higher the value of $p$, the more cases we have to consider, and  the calculation becomes more and more difficult. Nevertheless, we have been able to use this method for the second and third viriel coefficients ($b_2$ and $b_3$).
We notice immediatly that in order to calculate $n(\mu,T)$, hence $G({\bf p},0_{-})$, all the $e^{\mu\tau}$ terms in prefactors of the $G^{0}$'s ({\it cf.} Eq.\ref{eqG0}) cancel in the expressions. Therefore we do not consider them in the following. The chemical potential $\mu$ therefore appears only in the fugacity.

\begin{figure}[h]
\begin{center}
\includegraphics[width=0.5\linewidth]{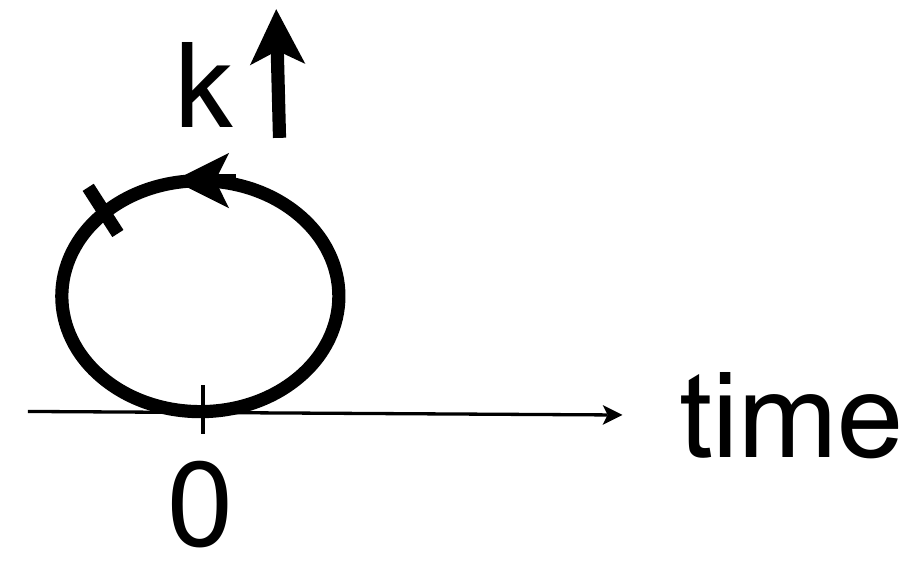}
\caption{The lowest order diagram.}
\label{figb1}
\end{center}
\end{figure}

At lowest order, we have simply $G\approx G^{(0,0)}+z\, G^{(0,1)}$. $G^{(0,0)}$ is retarded and therefore does not contribute to $G({\bf k},0_{-})$. Hence we find
\bea
\delta n^{(1)}&=& \sum_{{\bf k}}G^{(0,1)}({\bf k},\tau=0_{-})
\eea
which is shown in the diagram of Fig.\ref{figb1}. 
The calculation is easily done using Eq.\ref{eqg0nf} and we find
\bea
\delta n^{(1)}&=&(\Lambda_T)^{-3}
\eea
This is the ideal classical gas result.

In the following, we focus on the problem of distinguisable fermions of $2$ species $\uparrow$ (respectively $\downarrow$) of mass $m_{\uparrow}$ (respectively $m_{\downarrow}$) typical of the BEC-BCS crossover (for a review, see \cite{gps}).
Specifically, we consider equal masses ($m_{\uparrow}=m_{\downarrow}\equiv m$) and equal densities (unpolarized case) or equal chemical potential $\mu_{\uparrow}=\mu_{\downarrow}$. 
\section{Calculation of $b_2$}\label{b2}
\begin{figure}[h]
\subfigure[\label{figb2a}]{\includegraphics[width=0.49\linewidth]{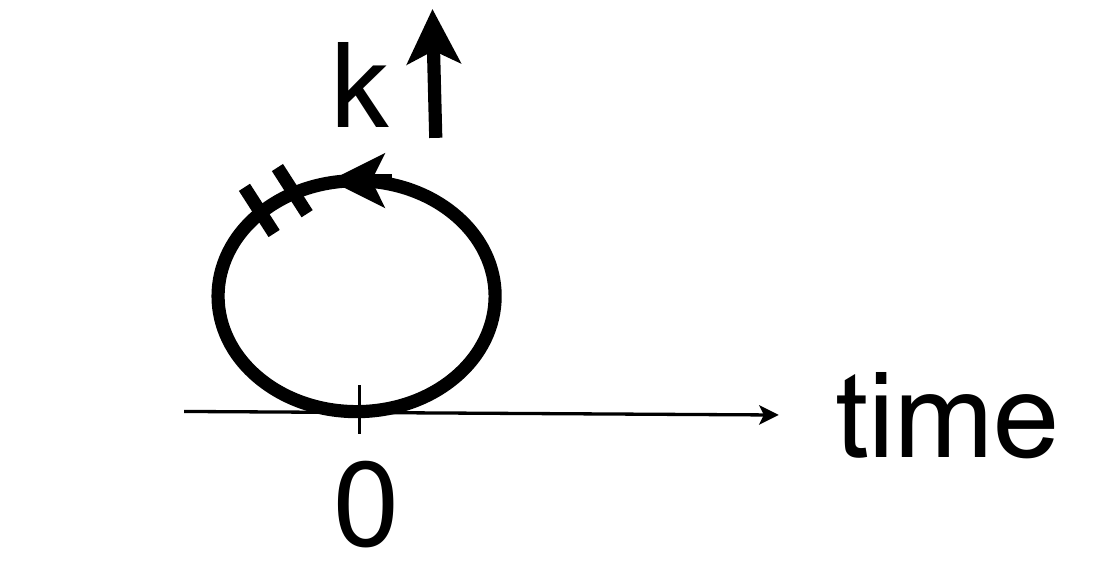}}\\
\subfigure[\label{figb2b}]{\includegraphics[width=0.65\linewidth]{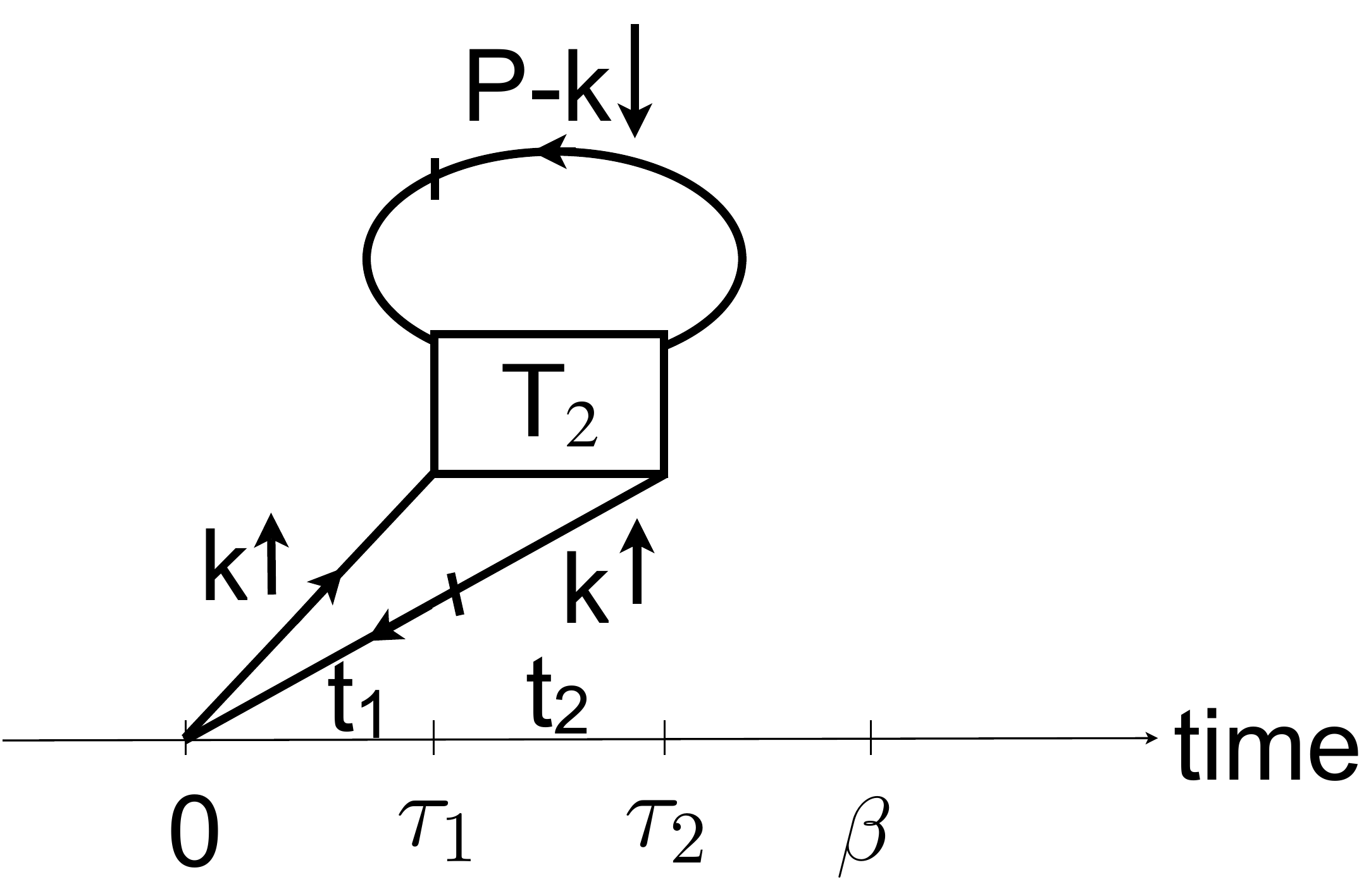}}
\caption{The two diagrams contributing to the virial coefficient $b_2$.
(a) is the diagrammatic expansion of the Fermi-Dirac distribution. (b) gives the Beth-Uhlenbeck contribution.\label{figb2}}
\end{figure}

In order to calculate $b_2$ \cite{vedelark}, we must consider all the diagrams with one $G^{(0,2)}$ (Fig.\ref{figb2a}) or $2$ $G^{(0,1)}$'s (Fig.\ref{figb2b}).

The two diagrams contributing to $\delta n^{(2)}$ are shown in Fig.\ref{figb2}.
The diagram of Fig.\ref{figb2a} gives the contribution 
\bea
\delta n^{(2,a)}&=&\sum_{{\bf k}}G^{(0,2)}({\bf k},\tau=0_{-})
\eea
One easily finds using Eqs.\ref{eqg0nf}
\bea
\delta n^{(2,a)}&=&-(\Lambda_T)^{-3}\,2^{-3/2}
\eea
This is of course just the free fermions contribution.
  
We consider now the contribution of the diagram of Fig.\ref{figb2}(b). It contains explicitely the effect of interactions, through the $2$-body $T$ matrix. We show now that it gives the Beth-Uhlenbeck contribution \cite{bethuhle}.The imaginary time integration is in the time
domain $\{0<\tau_1<\tau_2<\beta\}$, corresponding to the domain for time differences
$\{t_1>0,t_2>0,\beta-(t_1+t_2)>0\}$.
We therefore find
\bea
\delta n^{(2,b)}&=&\sum_{{\bf k},{\bf P}}\int dt_1\,dt_2\Theta(t_1)\Theta(t_2)
\Theta(\beta-(t_1+t_2))\nonumber\\
&&T_2({\bf P},t_2)e^{-(\beta-t_2)(\varepsilon_{\uparrow\, {\bf k}}+ \varepsilon_{\downarrow\, {\bf P- k}})}\label{eqt1t2n2b}
\eea
where $\varepsilon_{\uparrow,\downarrow\, {\bf k}}={\bf k}^2/(2m_{\uparrow,\downarrow})$.
The integration on $t_1$ is easily performed and gives a factor $(\beta-t_2)\Theta(\beta-t_2)$. The remaining integral on $t_2$ (for given ${\bf P}$ and ${\bf k}$) is of the form
\bea
I&=&\int dt_2 f_1(\beta-t_2) f_2(t_2)
\eea
with $f_1$ and $f_2$ two {\it retarded} functions ({\it i.e.} which are equal to zero for a non positive argument).
The convolution theorem for Laplace transforms gives
\bea
I&=&\int_{{\mathcal C}_{\gamma}}\frac{ds}{2\pi\,i}e^{-\beta s}F_1(s) F_2(s)\label{eqbetas}
\eea

${\mathcal C}_{\gamma}$ is the Bromwich contour of the problem. It is a straight line  in the $s$ complex plane parallel to the imaginary axis with a given real party $\gamma$. $\gamma$ is such that the integrand is analytic to the the left of the contour. We have also defined $F_{1}(s)=\int_{0}^{+\infty}d\tau e^{s \tau}f_1(\tau)$ (and similarly for $F_2(s)$)
for a complex variable $s$. The usual Laplace transform is defined with a minus sign in the exponential, but the  usual expressions for propagators are recovered with this definition.

In our problem, we have $f_1(\tau)=T_2({\bf P},\tau)$ and $f_2(\tau)=\tau e^{-\tau(\varepsilon_{\uparrow\, {\bf k}}+ \varepsilon_{\downarrow\, {\bf P- k}})}$. We easily find $F_2(s)=1/(s-(\varepsilon_{\uparrow\, {\bf k}}+ \varepsilon_{\downarrow\, {\bf P- k}}))^2$, and 
$F_1(s)=t_2(s-{\bf P}^2/(2 M))$, with 
\bea
M&=&m_{\uparrow}+m_{\downarrow}=2\,m\\
\eea
 the total mass of the $2$-body problem,
and (in $3D$) 
\bea
t_2(s)&=&\frac{2\pi}{m_r}\frac{1}{(a^{-1}-\sqrt{-2 m_r s})}
\eea This expression for $t_2(s)$ is easily found by writing down the integral (ladder) equation in time and taking the Laplace transform. This is just the usual result for a dimer propagator invacuum, with energy $s$. $a$ is the scattering length and 
\bea
m_r&=&\frac{m_{\uparrow}\,m_{\downarrow}}{(m_{\uparrow}+m_{\downarrow})}
=\frac{m}{2}
\eea
 the $2$-body reduced mass.
Next we go to the center of mass frame and make the change of variable
${\bf k}={\bf k}'+ (m_{\uparrow}/M){\bf P}$. By the change of variable $s'=s-{\bf P}^2/(2M)$, the center of mass momentum ${\bf P}$ decouples. The integration on ${\bf P}$ gives a prefactor
$(M/m_{\uparrow})^{3/2}(\Lambda_{\uparrow\, T})^{-3}$. Here $\Lambda_{\uparrow\, T}$ is the thermal wavelength
for $\uparrow$ particles. The integration on ${\bf k}'$ is also easily performed using
(in $3D$)
\bea
\sum_{{\bf k}'}\frac{1}{(s'-\frac{{\bf k}'^2}{2 m_r})^2}&=&\frac{(m_r)^2}{2\pi}\frac{1}{\sqrt{-2m_r s'}}\label{eqsum}
\eea
Finaly, we get the result
\bea
\delta n^{(2,b)}(\Lambda_{\uparrow\, T})^{3}&=&\left(\frac{M}{m_{\uparrow}}\right)^{3/2}
\int_{{\mathcal C}_{\gamma}} \frac{ds'}{2\pi\,i}e^{-\beta s'}\frac{t_2(s')}{\sqrt{-2m_r s'}}
\frac{m_r^2}{2\pi}\nonumber\\
&&\label{eqb2s}
\eea
We show now that we recover the Beth-Uhlenbeck result \cite{bethuhle} from Eq.(\ref{eqb2s}).
The integral on the variable $s'$ is transformed  by deforming the integration contour
along the real axis. There comes two kinds of contributions.
The first contribution comes from the molecular pole of $t_2(s)$ (if there is one) at the molecular energy $-|E_b|=-1/(2m_r a^2)$. The second contribution comes from the branch cut along the positive part of the real axis (physically the continuum of scattering states).We finally get for the contribution to the density 
\bea
\delta n^{(2,b)}\Lambda_{\uparrow T}^{3}&=&\left(\frac{M}{m_{\uparrow}}\right)^{3/2}\left\{
\bar{n}_{mol}+\bar{n}_{scatt.}\right\}
\eea
with
\bea
\bar{n}_{mol}&=&e^{\beta |E_b|},\mathrm{ if}\, a>0\label{eqnbarmol1}\\
&=&0,\qquad \mathrm{if}\,  a\leq 0\label{bethUhl1}
\eea
\bea
\bar{n}_{scatt.}&=&-\frac{1}{\pi}\int_{0}^{+\infty}\hspace{-.4cm}dx\,e^{-\beta\,x}
\frac{m_r}{\sqrt{2m_r x}}\frac{a^{-1}}{\left(a^{-2}+2 m_r x\right)}
\eea
We can recover the Beth-Uhlenbeck \cite{bethuhle} result by making the change of variable $x=p^2/(2m_r)$ 
\bea
\bar{n}_{scatt.}&=&-\frac{1}{\pi}\int_{0}^{+\infty}dp\,e^{-\beta\frac{p^2}{2 m_r}}
\frac{a}{\left(1+p^2 a^2\right)}\label{bethUhl2}
\eea

For equal masses, ($m_{\uparrow}=m_{\downarrow}$) we have for the second order virial coefficient $b_2$
\bea
b_2-2^{-5/2}&=&\sqrt{2}\left(\bar{n}_{mol}+\bar{n}_{scatt.}\right)
\eea
where $\bar{n}_{mol}$ is given by Eqs.(\ref{eqnbarmol1}),(\ref{bethUhl1}) and $\bar{n}_{scatt.}$
by Eq.(\ref{bethUhl2}).
In the unitary limit, if $a\to+\infty$ we find from Eqs(\ref{bethUhl1},\ref{bethUhl2}) $\bar{n}_{mol}=1$ and $\bar{n}_{scatt.}=-1/2$ thus (for equal masses $m_{\downarrow}=m_{\uparrow}$) $\delta n^{(2,b)}\Lambda_{T}^{3}=\sqrt{2}$ (corresponding to $b_2=1/\sqrt{2}$ \cite{muelho}). If $a\to-\infty$, $\bar{n}_{mol}=0$ and $\bar{n}_{scatt.}=1/2$ and we find the same result.
\section{Calculation of $b_3$}\label{b3}
We now come  to the calculation of the third virial coefficient $b_3$, which is the main result of this work. We focus on the $3D$ situation but the method works in any dimension. We proceed following the general method presented in section \ref{general} and developped in section \ref{b2}. We want to gather all the diagrams of order $z^3$.
There are three sorts of diagrams (the partition number of $3$) : 1) with one $G^{(0,3)}$, 
2) with one $G^{(0,2)}$ and one $G^{(0,1)}$, 3) with three $G^{(0,1)}$'s.
We will denote the corresponding corrections to the density
$\delta n^{(3,1)}$, $\delta n^{(3,2,\cdots)}$ and $\delta n^{(3,3,\cdots)}$ (there are several contributions in the last two cases).

The diagram 1) is nothing but the free fermion result and it amounts to expand the Fermi-Dirac distribution in power of $z$. This is a standard result, and we find for the density the correction for free fermions (in $d=3$ dimensions)
\bea
\delta n^{(3,1)}\Lambda_{T} ^3&=&3^{-3/2}
\eea
\subsection{$\delta n^{(3,2)}$}
\begin{figure}[h]
\subfigure[\label{figb32a}]{\includegraphics[width=0.49\linewidth]{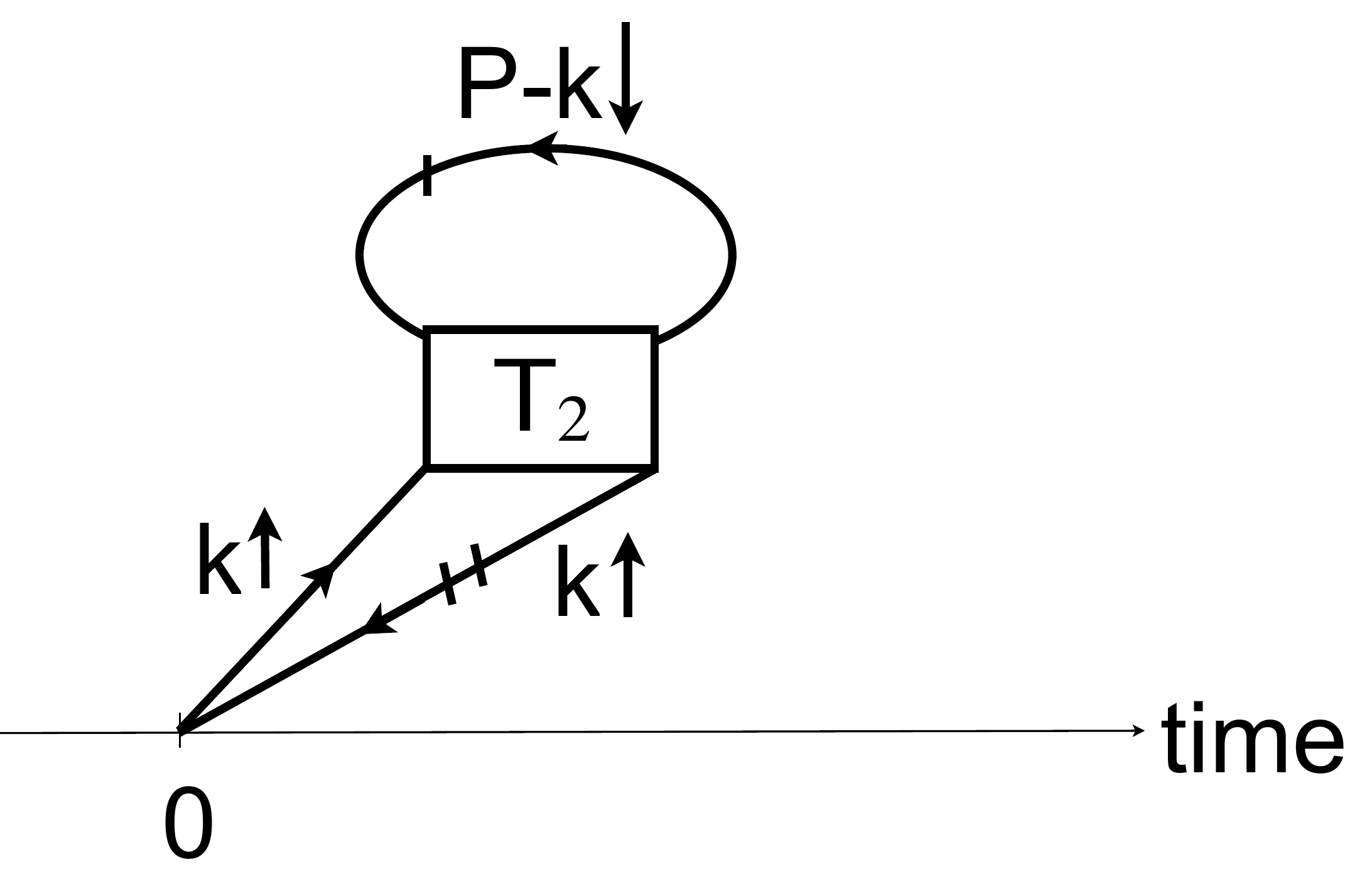}}
\subfigure[\label{figb32b}]{\includegraphics[width=0.49\linewidth]{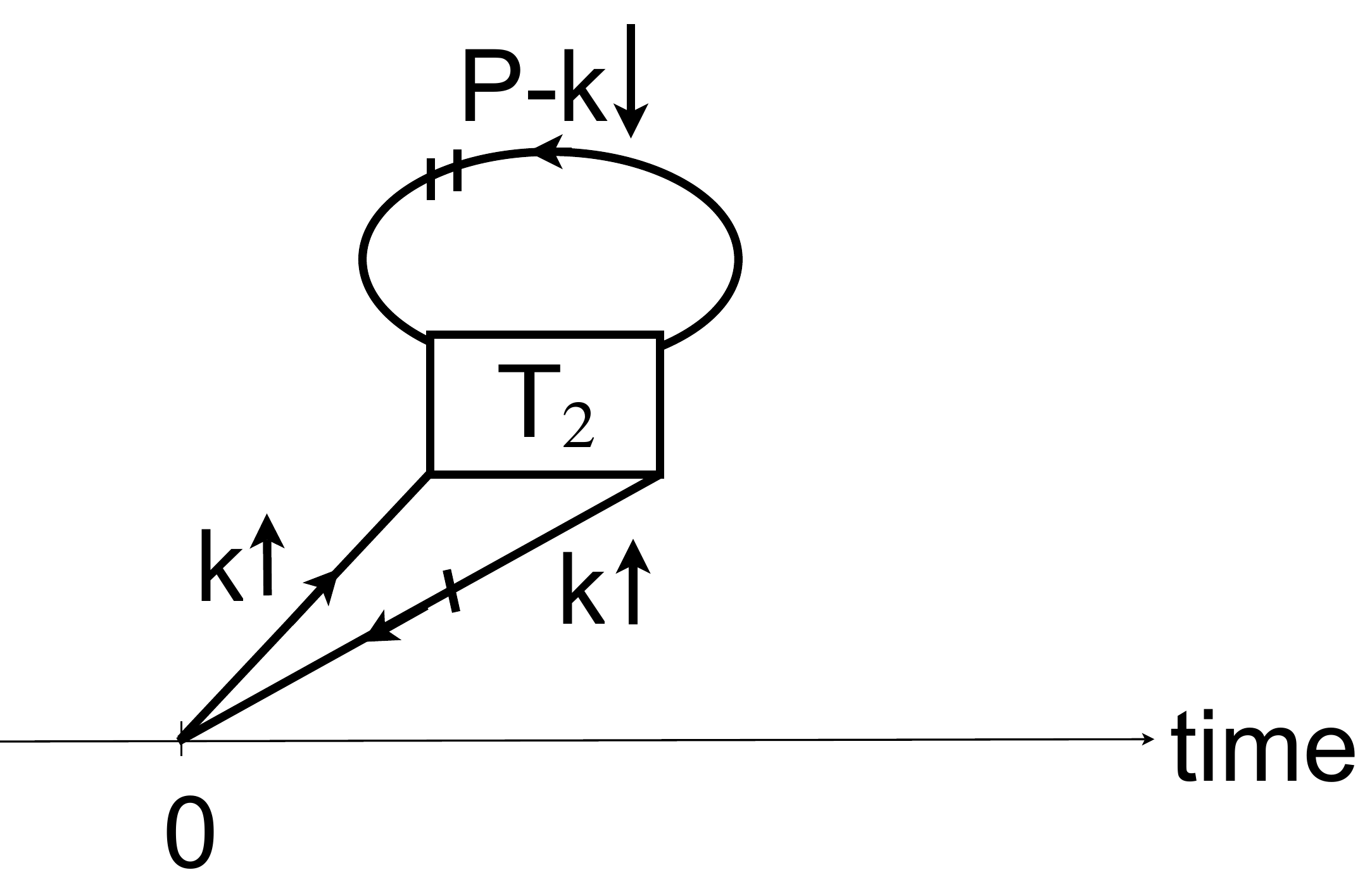}}
\caption{The two diagrams contributing to $b_3$ with one $G^{(0,2)}$ and one $G^{(0,1)}$ (see text). \label{figb32}}
\end{figure}

The diagrams with one $G^{(0,2)}$ and one $G^{(0,1)}$ are shown in Fig.\ref{figb32}.
The diagrams of Fig.\ref{figb32a} and \ref{figb32b} are very similar to the diagram of Fig.\ref{figb2}(b). Indeed, by noticing that $G^{(0,2)}({\bf k},-(t_1+t_2))=(-e^{-\beta\varepsilon_{{\bf k}}})G^{(0,1)}({\bf k},-(t_1+t_2))$ we just have  to multiply
the time integral in Eq.(\ref{eqt1t2n2b}) by $-e^{-\beta\varepsilon_{\uparrow{\bf k}}}$ in order to get the contribution of diagram of Fig.\ref{figb32}(a).  We can therefore skip directly to Eq.(\ref{eqbetas}) in order to evaluate the time integrals. Then we make the following change of variables (the Jacobian is unity)
\bea
s'&=&s-\frac{{\bf P}^2}{4\,m}\\
{\bf k}'&=&{\bf k}-\frac{1}{2}{\bf P}\\
{\bf P}'&=&\frac{2}{3} \left({\bf k}
+{\bf P}\right)
\eea
The integral on ${\bf P}'$ decouples and can be done analytically. The angular integration on ${\bf k}'$ can also
be performed, and we get
\begin{widetext}
\bea
\delta n^{(3,2,a)}\Lambda_{\uparrow T}^3&=&
\frac{8}{3\sqrt{3}}
\int_0^{+\infty}\frac{dk'\,k'^2}{2\pi^2} e^{-\beta\frac{k^2}{3\, m}}
\int_{\mathcal{C}}
\frac{ds'}{2\pi\,i}e^{-\beta s'}\,
\frac{t_2(s')}{
(s'-\frac{k^2}{2\,m_r})^2
}\label{eqres32a}
\eea
\end{widetext}

The diagram of Fig.\ref{figb32b} can be calculated along the same line. We just have to exchange the role of $\uparrow$ and $\downarrow$ particles. As a result we find
\bea
\delta n^{(3,2,b)}&=&\delta n^{(3,2,a)}
\eea

At unitarity ($a^{-1}=0$), the integrals can be done analytically. 
\bea
\delta n^{(3,2,a)}\Lambda_{T}^3&=&\delta n^{(3,2,b)}\Lambda_{T}^3\nonumber\\
&=&-\left(\frac{8}{9\sqrt{3}}-\frac{2}{3\pi}\right)\approx-0.300994\label{eq331lu}
\eea
\subsection{$\delta n^{(3,3)}$}
We want to find all the diagrams with three $G^{(0,1)}$ propagators. A given diagram always contains an incoming ${\bf k},\uparrow$ line to the imaginary time $\tau=0$.
The initial time of this fermionic line must be in the interval $[0,\beta]$. Therefore, this line can not correspond to a retarded propagator $G^{(0,0)}$, and it must be a $G^{(0,1)}_{\uparrow}$ (slashed) line. A given diagram also always contains an outgoing ${\bf k},\uparrow$ line from the imaginary time $\tau=0$. This can be a $G^{(0,1)}$ (slashed) line ({\it first case}) or a retarded $G^{(0,0)}$ line ({\it second case}).
\subsubsection{$\delta n^{(3,3,1)}$}
\begin{figure}[h]
\begin{center}
\includegraphics[width=.7\linewidth]{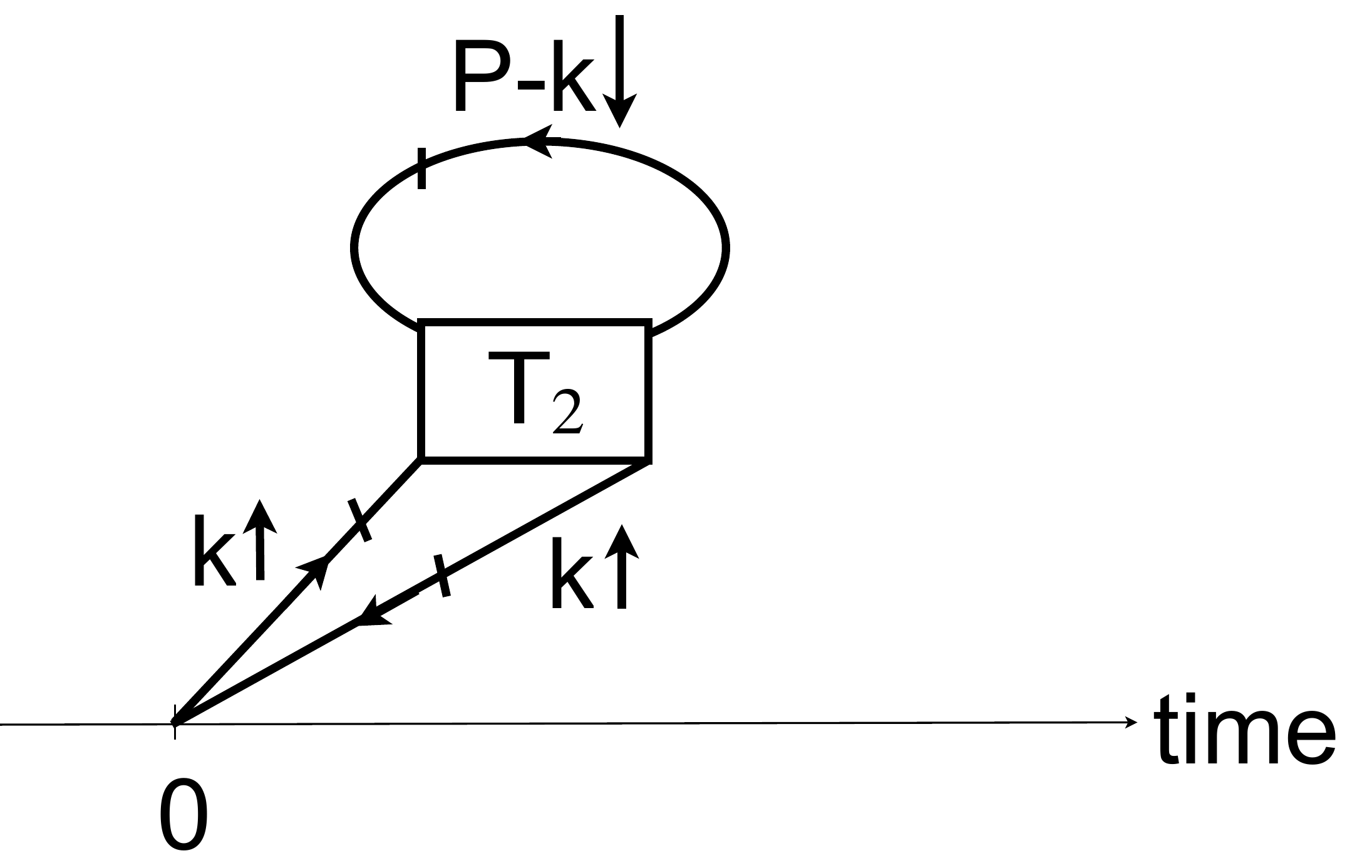}
\caption{The diagram contributing to $\delta n^{(3,3,1)}$}
\label{figb331}
\end{center}
\end{figure}

In the first case, the only possible diagram is shown in Fig.\ref{figb331} and is similar to diagrams of Figs.\ref{figb2b}, and Fig.\ref{figb32}.
If we follow the same line of reasoning than for the calculation of these diagrams, we easily find \bea
\delta n^{(3,3,1)}&=&\delta n^{(3,2,a)}
\eea
\begin{figure}[h]
\begin{center}
\includegraphics[width=8cm]{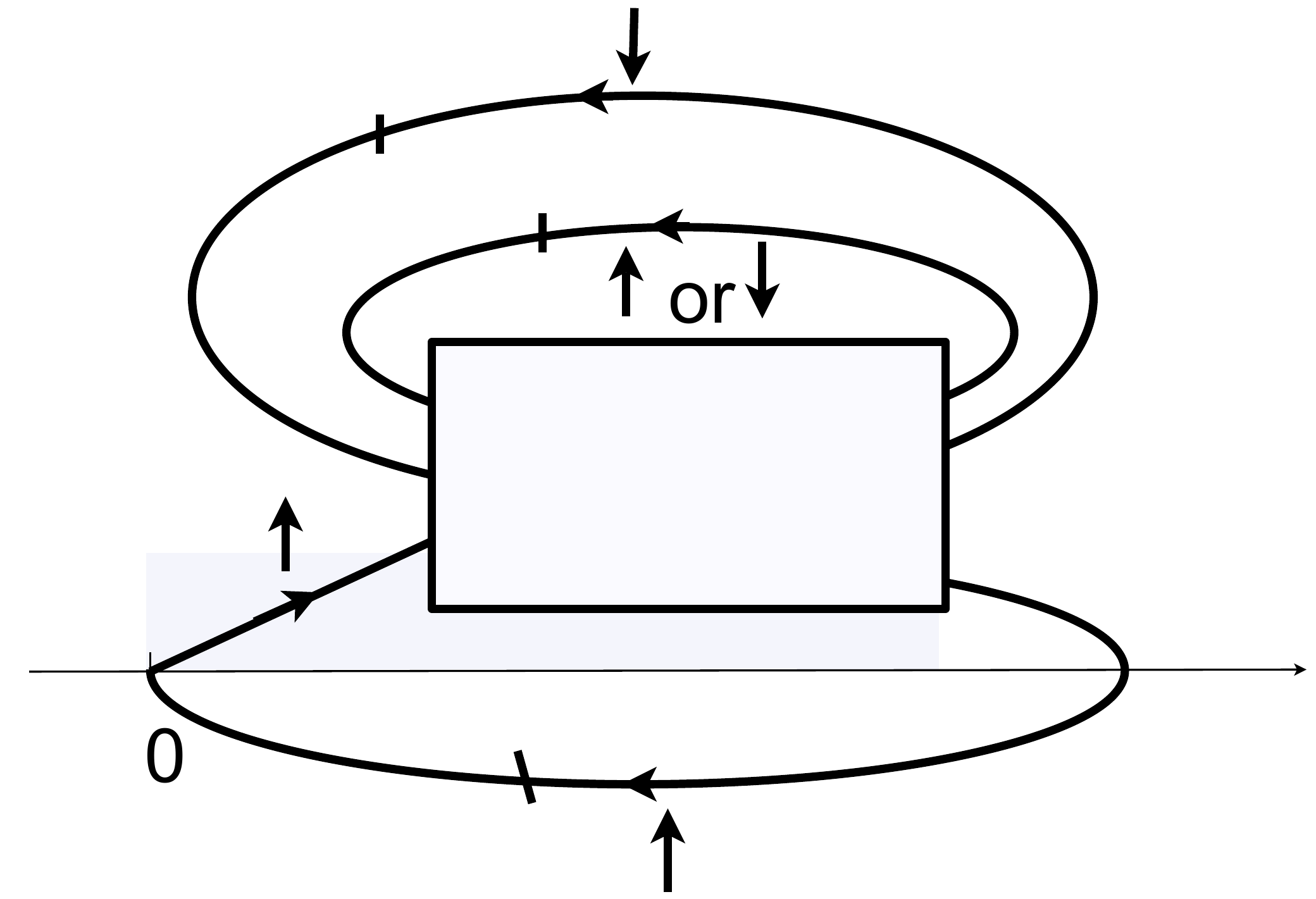}
\caption{The general structure of a diagram contributing to $\delta n^{(3,3,2)}$}
\label{figb33gen}
\end{center}
\end{figure}

\subsubsection{$\delta n^{(3,3,2)}$}
\begin{figure}[h]
\subfigure[\label{figb332a}]{\includegraphics[width=0.49\linewidth]{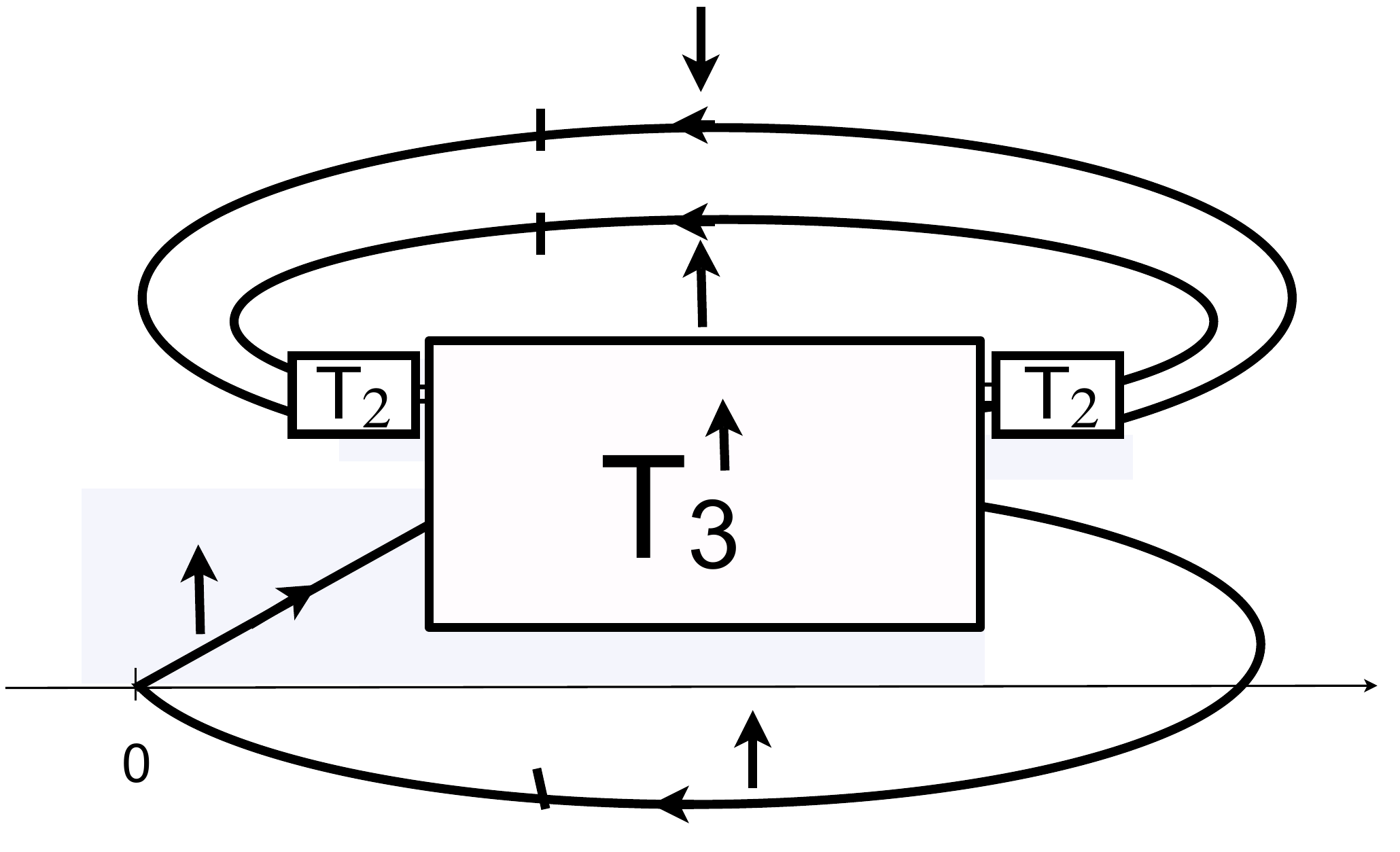}}
\subfigure[\label{figb332b}]{\includegraphics[width=0.49\linewidth]{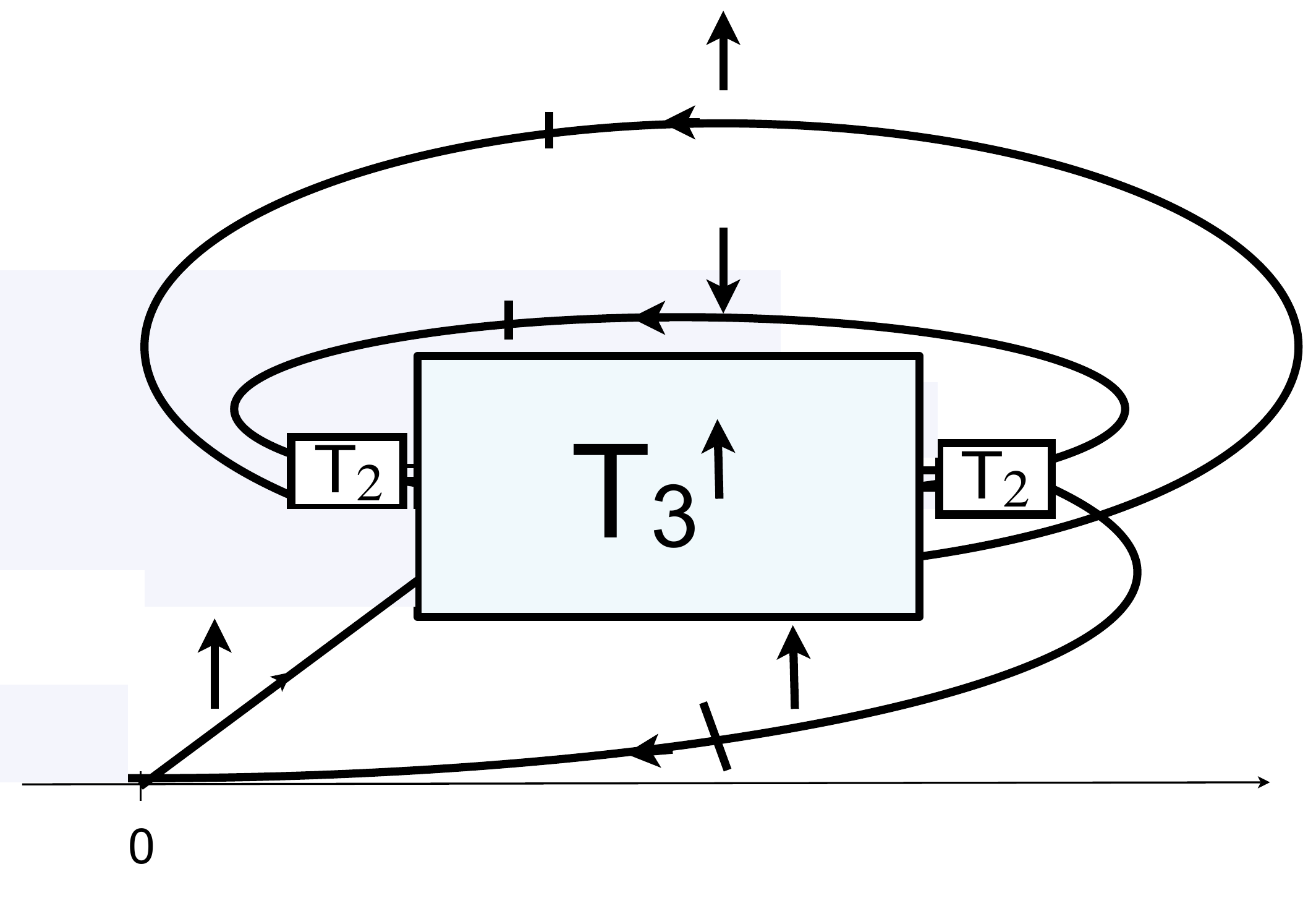}}\\
\subfigure[\label{figb332c}]{\includegraphics[width=0.48\linewidth]{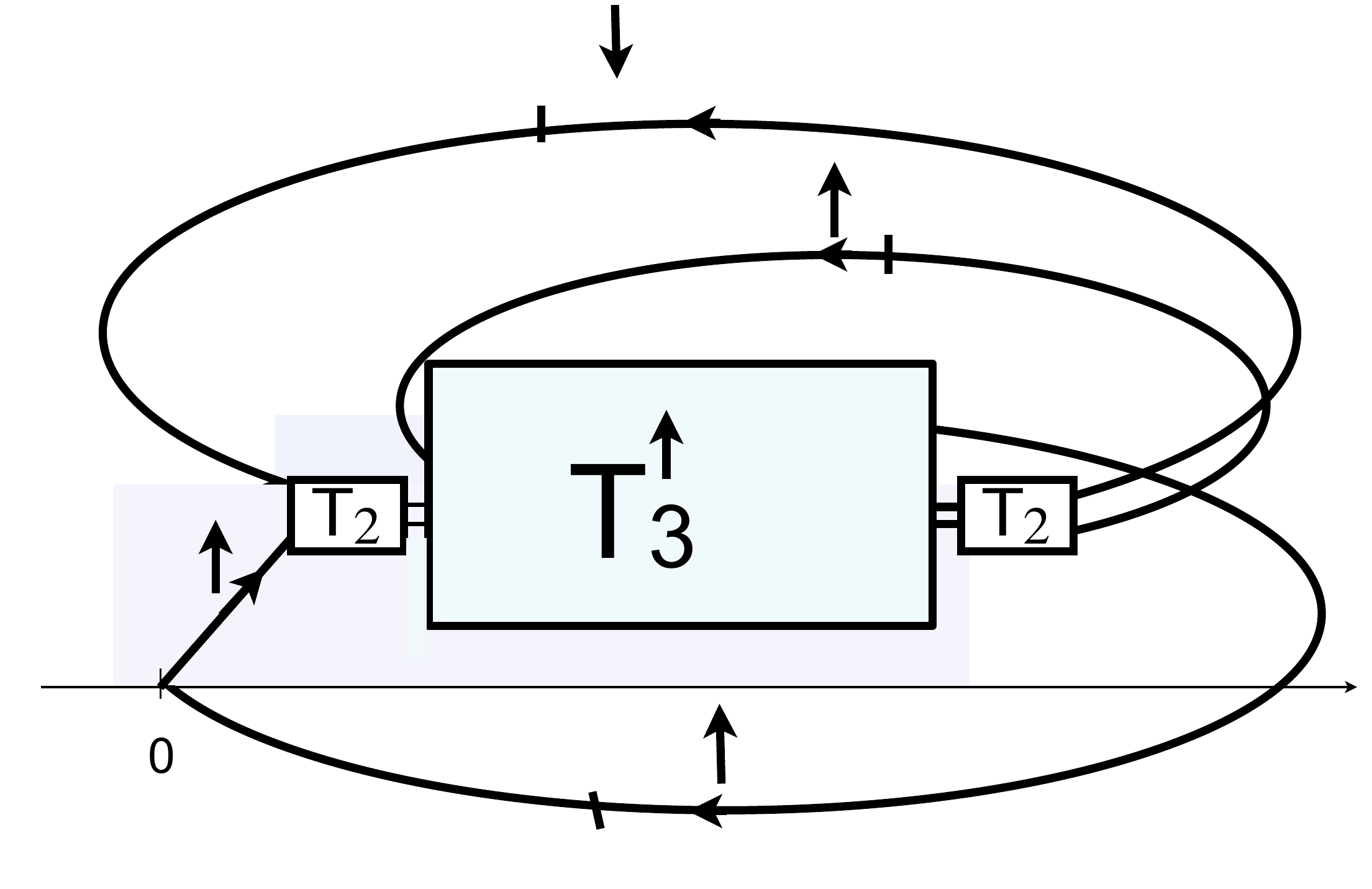}}
\subfigure[\label{figb332d}]{\includegraphics[width=0.48\linewidth]{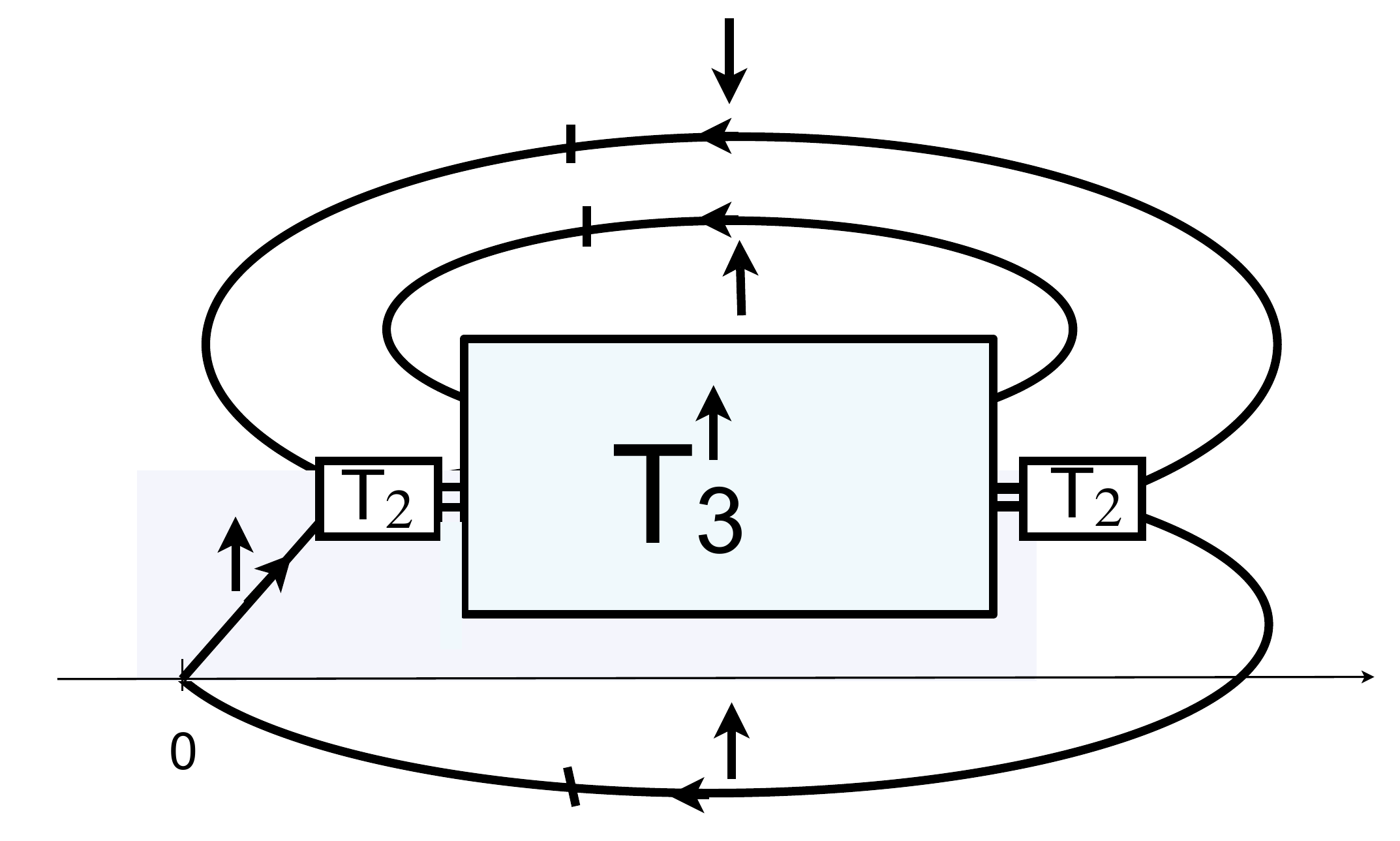}}\\
\subfigure[\label{figb332e}]{\includegraphics[width=0.48\linewidth]{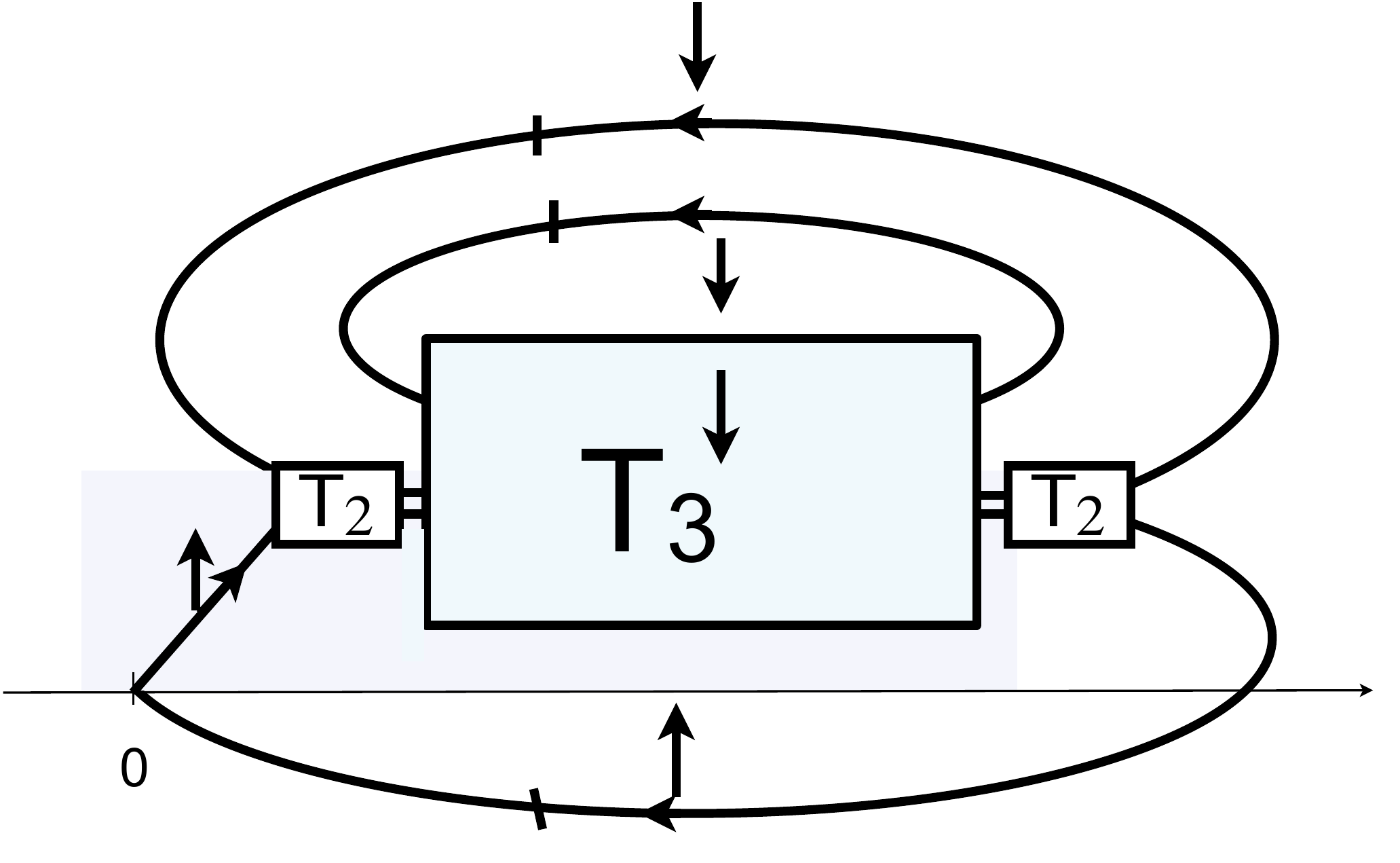}}
\subfigure[\label{figb332f}]{\includegraphics[width=0.48\linewidth]{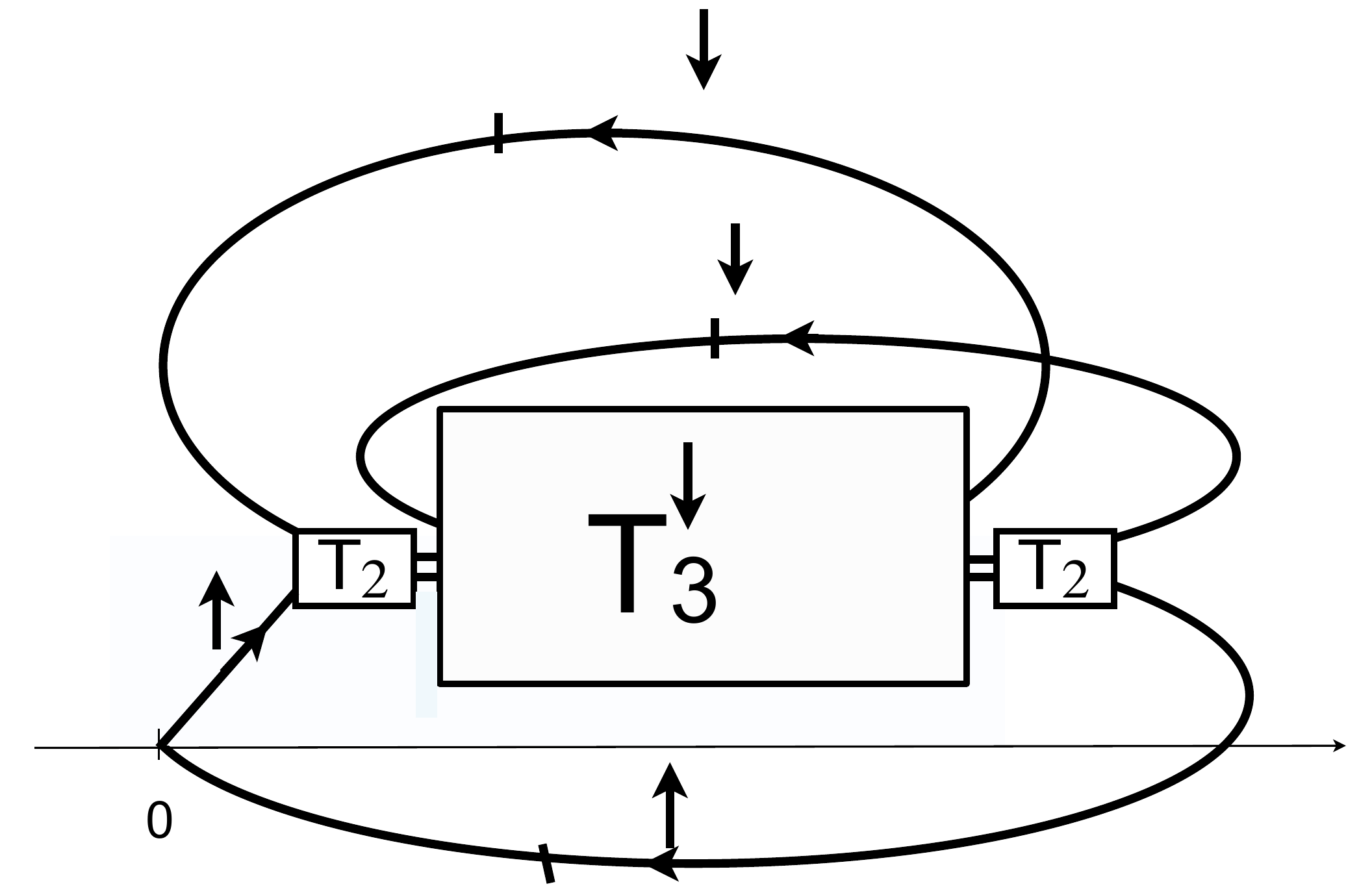}}
\caption{The  $6$ diagrams contributing to $\delta n^{(3,3,2)}$ (see text).}
\label{figb332af}
\end{figure}

The second case is the most interesting. A general diagram is drawn in Fig.\ref{figb33gen}. It contains two $G^{(0,1)}_{\uparrow}$'s and one  $G^{(0,1)}_{\downarrow}$ (Fig.\ref{figb332a} to \ref{figb332d}) or
two $G^{(0,1)}_{\downarrow}$'s and one  $G^{(0,1)}_{\uparrow}$ (Fig.\ref{figb332e} and \ref{figb332f}).

We consider first the case with two $G^{(0,1)}_{\uparrow}$'s and one  $G^{(0,1)}_{\downarrow}$.
The incoming $\downarrow$ (from the left) can first interact with the incoming (from the left) $G^{(0,1)}_{\uparrow}$
(diagrams of Fig.\ref{figb332a} and \ref{figb332b}) or with the incoming $G^{(0,0)}_{\uparrow}$ (diagrams of Fig.\ref{figb332c} and \ref{figb332d}). In each case, the two diagrams are found by choosing which line $G^{(0,1)}_{\uparrow}$ the outgoing (to the right) $\downarrow$ can last interact with. In all these diagrams, 
$T_3^{\uparrow}$ is the $3$-body $T$-matrix, extensively used in \cite{pra4par,prllhy,prallhypol}. It physically describes the scattering 
of a $\uparrow$ particle with a dimer (diagrammatically represented by $T_2$).

The case with two $G^{(0,1)}_{\downarrow}$'s and one  $G^{(0,1)}_{\uparrow}$ can be treated in a similar way : in
Fig.\ref{figb332e}, the external $\uparrow$ line interact with the same $G^{(0,1)}_{\downarrow}$ line at the entrance (on the left) and at the exit (on the right) of $T_3^{\downarrow}$; whereas in Fig.\ref{figb332f}, the two
$G^{(0,1)}_{\downarrow}$'s have been exchanged at the exit of $T_3^{\downarrow}$.

We first consider the calculation of $\delta n^{(3,3,2,a)}$. This is explained in Appendix \ref{app332a}. The difficulty is to perform the time integrals. It is however possible
to transform a multidimensional time integral into a simple integration on a (complex) energy. Mathematically, this comes from the fact that the time integrals are just convolutions, and the convolution theorem for Laplace transforms enables to write it as an inverse Laplace transforms of products of functions of the energy, as we did for the calculation of $b_2$.
We therefore give the results 
\begin{widetext}
\bea
\delta n^{(3,3,2,a)}\Lambda_{T}^3&=&-3^{3/2}\frac{m_r^2}{2\pi}
\int_{\mathcal{C}}\frac{ds'}{2\pi\,i}e^{-\beta s'}\, \int_0^{+\infty}\frac{dp\,p^2}{2\pi^2}
\frac{\left[t_2(s'-\frac{p^2}{2\,m_{A\uparrow,D}})\right]^2}{
\sqrt{-2\,m_r(s'-\frac{p^2}{2\,m_{A\uparrow,D}})}
}
\left(\sum_{l\geq 0} t_{3,l}^{\uparrow}( p,p;s')\right)\label{eqres332a}
\eea
\end{widetext}
$\mathcal{C}$ is a Bromwhich contour
such that the integrand is analytical to the right of $\mathcal{C}$. 
\bea
m_{A\uparrow,D}&=&\frac{m_{\uparrow}(m_{\uparrow}+m_{\downarrow})}{(2 m_{\uparrow}+m_{\downarrow})}=\frac{2}{3}m
\eea is the atom$\uparrow$-dimer reduced mass.

The calculation of $\delta n^{(3,3,2,b)}$ is similar, and we just give the result
\begin{widetext}
\bea
\delta n^{(3,3,2,b)}\Lambda_{T}^3&=&3^{3/2}
\sum_{l \geq 0}
\int_{\mathcal{C}}
\frac{ds'}{2\pi\,i}e^{-\beta s'}
\int_0^{+\infty}\frac{dp}{2\pi^2}p^2 
 t_2(s'-\frac{p^2}{2\,m_{A\uparrow,D}})
\int_0^{+\infty}\frac{dp'}{2\pi^2}p'^2 
 t_2(s'-\frac{p'^2}{2\,m_{A\uparrow,D}})\nonumber\\
 &&\times
\left(  \frac{m}{p\,p'}  \right)^2
\tilde{Q}_l\left( 
\frac{m}{p\,p'}\left(s'-\frac{p^2+p'^2}{2\,m_r}\right)
\right)
 t_{3,l}^{\uparrow}( p, p';s')\label{eqres332b}
\eea
\end{widetext}
with $\tilde{Q}_l(z)\equiv -\frac{d Q_l(z)}{d z}=\frac{1}{2}\int_{-1}^{1} du \frac{P_l(u)}{(z-u)^2}$ ($Q_l$ is the Legendre function of the second kind).

The calculation of diagrams of Fig.\ref{figb332c} and \ref{figb332d} is done in the same way and we find that
\bea
\delta n^{(3,3,2,d)}&=&\delta n^{(3,3,2,a)}\\
\delta n^{(3,3,2,c)}&=&\delta n^{(3,3,2,b)}
\eea
Moreover, it is also easy to verify that $\delta n^{(3,3,2,e)}$ (respectively $\delta n^{(3,3,2,f)}$) is obtained from the expression for $\delta n^{(3,3,2,a)}$ (respectively $\delta n^{(3,3,2,b)}$) by exchanging the role of $\uparrow$ and $\downarrow$ particles (this amounts to exchange $\mup$ and $\md$ in the analytical expressions).

\subsection{Calculation of $b_3$ in the case $\mup=\md$}
\begin{figure}[h]
\begin{center}
\includegraphics[width=8cm]{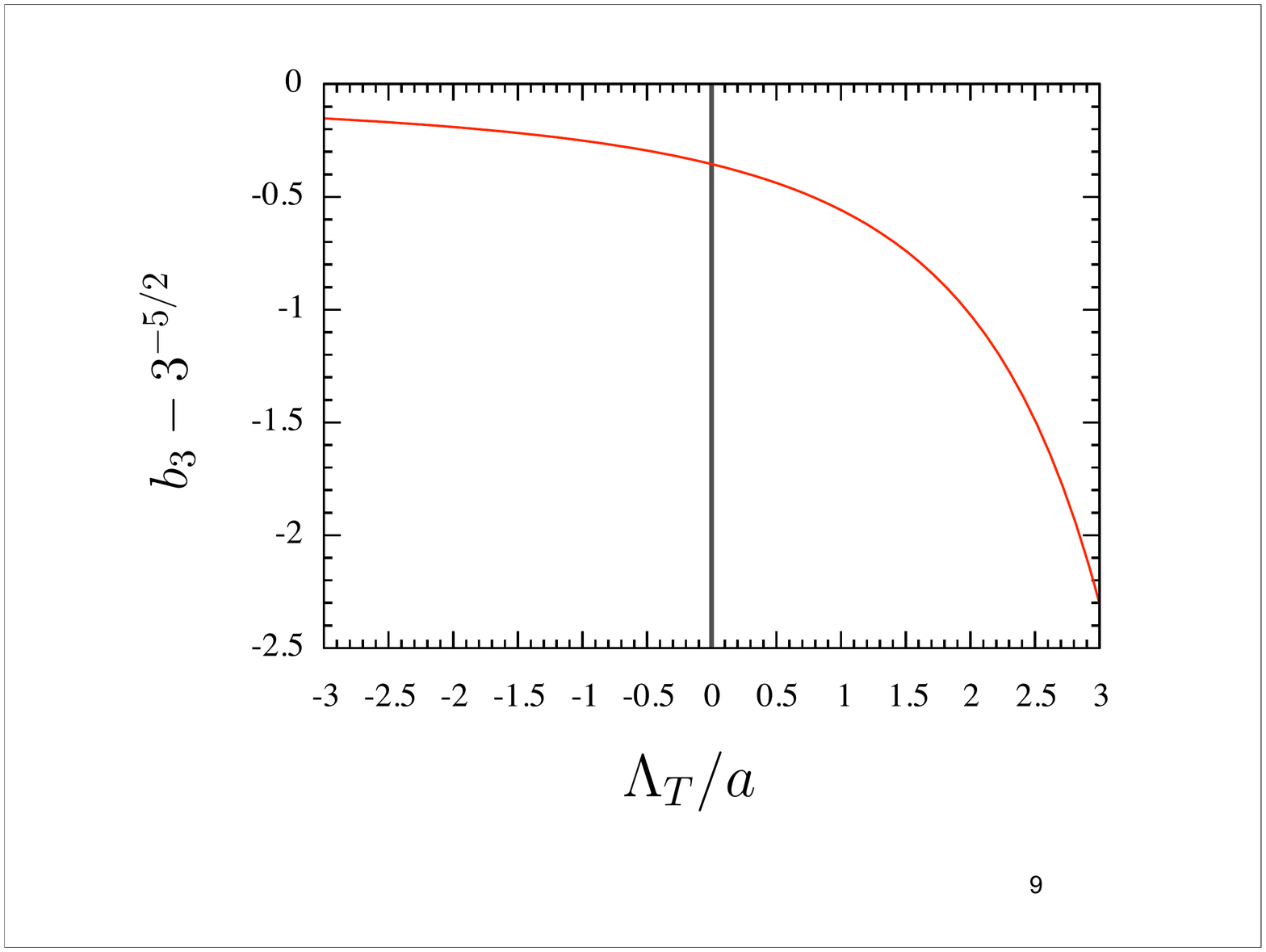}
\caption{The third virial coefficient $b_3-3^{-5/2}$ as a function of the dimensionless parameter $\Lambda_T/a$.}
\label{figb3}
\end{center}
\end{figure}

In general, $b_3$ is obtained from $\delta n^{(3)}$ by summing all the contributions
$\delta n^{(3,1)}, \delta n^{(3,2,a\,{\mathrm and}\,b)},\delta n^{(3,3,1)}$, and the six $\delta n^{(3,3,2)}$'s.
However in the case of equal masses, we have $\delta n^{(3,3,1)}=\delta n^{(3,2,a)}=\delta n^{(3,2,b)}$ , $\delta n^{(3,3,2,a)}=\delta n^{(3,3,2,d)}=\delta n^{(3,3,2,e)}$ ,
In this way we get
\bea
\delta n^{(3)}\Lambda_T^3&=&3^{-3/2}+3  \left(\delta n^{(3,2,a)}
+\delta n^{(3,3,2,a)}+\delta n^{(3,3,2,b)}
\right)\Lambda_T^3\nonumber\\
&&
\eea
And therefore for $b_3=\delta n^{(3)}\Lambda_T^3 /3$ 
\bea
b_3&=&3^{-5/2}+\left(\delta n^{(3,2,a)}
+\delta n^{(3,3,2,a)}+\delta n^{(3,3,2,b)}
\right)\Lambda_T^3 \nonumber\\
&&\label{eqb3fin}
\eea
The numerical result is shown in Fig.\ref{figb3}. In order to reveal the effect of interactions, we have plotted the difference $b_3-3^{-5/2}$ as a function of $\Lambda_T/a$. In practice, in order to integrate safely on $s'$, we deform the contour ${\mathcal C}$ into two semi straight lines, symetric with respect to the real axis, and making an angle $\alpha\in ]0,\pi/2[$
with it. In this way, the exponential goes to zero at the extremities of the contour. We checked for different values of $\alpha$, that the result is independant of $\alpha$.
We discretized the values of $s'$, $p$ and $p'$ on a grid in order to determine
$t^{\uparrow}_{3,l}$ numerically from the linear integral equation (\ref{eqt3l}), which transforms into the problem of solving linear algebraic equations. Finally, for a given value of $l$, we sum on $s'$,$p$ and $p'$ in order to get $\delta n^{(3,3,2,a)}$ and $\delta n^{(3,3,2,b)}$.
The curve of Fig.\ref{figb3} is in excellent agreement with the results of Ref.\cite{hld}.
At unitarity, we find for the $b_3$ coefficient (we sum the harmonics in Eqs.(\ref{eqres332a}),(\ref{eqres332b}) up to $l=7$ for the four digits we show)
\bea
b_3-3^{-5/2}&\approx&-0.3551\label{eqb3lunum}
\eea
in excellent agreement with the results of Refs\cite{hld,rdblume}. This is therefore an independent check for the value of $b_3$, since our method is very different from the method of \cite{hld,rdblume}.

\section{Conclusion}\label{conclusion}

We have calculated the third order virial coefficient $b_3$ in the problem of $2$-species fermions interacting via a short range interaction.  We have developped a diagrammatic method which leads to explicit analytic expressions. This approach might be extended to the calculation of the $4$th order virial coefficient $b_4$, which was recently calculated
\cite{rdblume}.

{\bf Acknowledgments:} We thank R. Combescot, F. Chevy, C. Mora, N. Navon, N. Regnault and F. Werner
for stimulating discussions.
\appendix\label{details}
\section{Calculation of $\delta n^{(3,3,2,a)}$}\label{app332a}
\begin{figure}[h]
\includegraphics[width=8.cm]{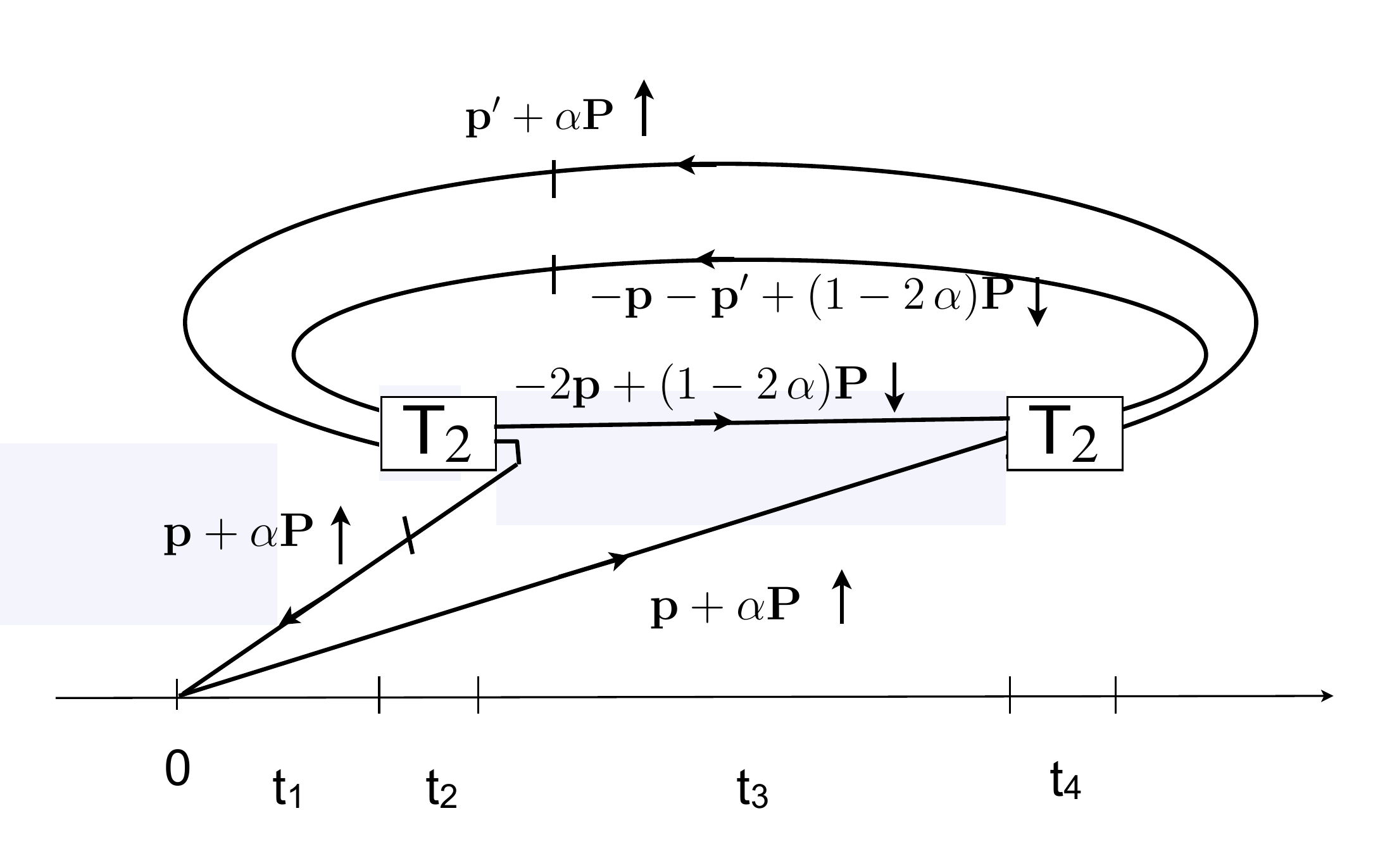}
\caption{Diagram of Fig.\ref{figb332a}, with $T_3^{\uparrow}$ taken in the Born approximation.}
\label{figb332aborn}
\end{figure}

In order to calculate the diagram of Fig.\ref{figb332a}, we first look at the "Born" approximation of $T_3^{\uparrow}$. The diagram is shown in Fig.\ref{figb332aborn}. In this figure, we have indicated the $4$ relevant time differences $t_{1,\cdots 4}$ as well
as a convenient way to choose the momentum variables. Physically ${\bf P}$ is the total momentum of the $3$-body problem ($2\,\uparrow$, $1\,\downarrow$). ${\bf p}$ is the momentum of the incoming $\uparrow$, in the centre of mass reference frame, scattering with a dimer of opposite momentum in the centre of mass reference frame. 
We have defined
\bea
\alpha&=&\frac{\mup}{(2\,\mup+\md)}=\frac{1}{3}
\eea
We give expressions valid for any masses $\md$ and $\mup$.\\

The time domain is terms of time differences is
$
\{t_1>0,\cdots\,,t_4>0,\,\beta-\left( t_1+\cdots+t_4\right)>0   \}
$.
First, we notice that the integrand is independent of $t_1$. Integrating on $t_1$ gives
a factor $\left(\beta-(t_2+t_3+t_4)\right)\Theta\left(\beta-(t_2+t_3+t_4)\right)$.
After integration on $t_1$, we can then write this diagram as (we do not sum on momenta, and denote it $I$)
\begin{widetext}
\bea
I&=&-\int \Pi^4_{i=2}dt_{i}\,
\left(T_2\left(-{\bf p}+(1-\alpha){\bf P},t_2\right)
e^{-\varepsilon_{\uparrow}({\bf p}+\alpha{\bf P})t_2}\Theta(t_2)\right)
\left(
e^{-\left(
2\varepsilon_{\uparrow}({\bf p}+\alpha{\bf P})+
\varepsilon_{\downarrow}(-2{\bf p}+(1-2\alpha){\bf P})
\right)t_3
}\Theta(t_3)
\right)\nonumber
\\
&&
\times
\left(T_2\left(-{\bf p}'+(1-\alpha){\bf P},t_4\right)
e^{-\varepsilon_{\uparrow}({\bf p}'+\alpha{\bf P})t_4}\Theta(t_4)\right)\nonumber\\
&&\times
\left(
\Big(\beta-(t_2+t_3+t_4)\right)\Theta\left(\beta-(t_2+t_3+t_4)\right)
e^{-\left(\beta-(t_2+t_3+t_4)\right)\left(
\varepsilon_{\uparrow}({\bf p}+\alpha{\bf P})
+\varepsilon_{\uparrow}({\bf p}'+\alpha{\bf P})
+\varepsilon_{\downarrow}(-{\bf p}-{\bf p}'+(1-2\alpha){\bf P})
\right)} \Big)
\eea
\end{widetext}

This apparently complicated expression is actually a convolution integral of the type
studied in Appendix \ref{convol}. We then easily find
\begin{widetext}
\bea
I&=&\int_{{\mathcal C}}\frac{ds}{2\pi i}e^{-\beta\,s}
\left[
t_2\left(
s-\frac{{\bf P}^2}{2(2\mup+\md)}-\frac{{\bf p}^2}{2\,m_{A\uparrow,D}}
\right)
\right]^2
\frac{1}{\left(
(s-\frac{{\bf P}^2}{2(2\mup+\md)})-
(\frac{p^2+p'^2}{2 m_r}+\frac{{\bf p}\cdot{\bf p}'}{\md}
)\right)^2}\nonumber\\
&&\times
\frac{1}{\left(
(s-\frac{{\bf P}^2}{2(2\mup+\md)})-
\frac{p^2}{2}(\frac{4}{\mup}+\frac{2}{\md})
\right)}\label{eqI}
\eea
\end{widetext}
If we look at the next order Feynman diagrams for $T_3^{\uparrow}$, we see by inspection that we still find a convolution integral, and we just have to replace the last
line of the above equation by $t_{3}^{\uparrow}({\bf p},{\bf p};(s-\frac{{\bf P}^2}{2(2\mup+\md)}))$.  In this way, the multimensional time integrals are replaced by a single integral on the variable $s$.
 
We now come to the summation on wave vectors. For the summation on the total momentum ${\bf P}$, we see that, similarly to what was found in the calculation of $b_2$, ${\bf P}$ enters only through the combination $s-\frac{{\bf P}^2}{2(2\mup+\md)}$. This enables to decouple the summation on ${\bf P}$ (by the change of variable $s'=s-\frac{{\bf P}^2}{2(2\mup+\md)})$. After summation on ${\bf P}$, we get a factor
$(\Lambda_{\uparrow T})^{-3}\left(\frac{2\mup+\md}{\mup}\right)^{3/2}$.

Finally, for $\delta n^{(3,3,2,a)}$, it is also possible to integrate on ${\bf p}'$, since ${\bf p}'$ enters only the expression through
$
\frac{1}{\left(
(s-\frac{{\bf P}^2}{2(2\mup+\md)})-
(\frac{p^2+p'^2}{2 m_r}+\frac{{\bf p}\dot{\bf p}'}{\md}
)\right)^2}
$. The integration is done via the change of variable
${\bf p}''={\bf p}'+  (\mup)/(\mup+\md){\bf p}$, and the use of Eq.(\ref{eqsum}). In this way we get a factor
$(m_r)^2/(2\pi)/\sqrt{-2\,m_r(s'-p^2/(2\,m_{A\uparrow,D})}$.

The last step is to express $t_3^{\uparrow}({\bf p},{\bf p};s')$ in terms of the Legendre Polynomials components $t^{\uparrow}_{3,l}$'s. This is done thanks to
\bea
t_3^{\uparrow}({\bf p},{\bf p};s')&=&\sum_{l\geq 0}t^{\uparrow}_{3,l}(p,p;s')
\eea
which physically represents the {\it forward} scattering of an atom on a dimer (we used $P_l(1)=1$).
In this way, we recover Eq.(\ref{eqres332a}).

\section{Integral equation for the $3$-body problem in the center of mass frame of reference.}\label{appeqt3}
We first recall the integral equation for $T_3^{\uparrow}(p_1,p_2;P)$ (for complex frequencies) following the approach of \cite{pra4par}
\bea
T_3^{\uparrow}(p_1,p_2;P)&=&-G_{\downarrow}(P-p_1-p_2)\nonumber\\
&&
\hspace{-2.cm}
-\sum_q
G_{\uparrow}(q)G_{\downarrow}(P-p_1-q)T_2(P-q)T_3^{\uparrow}(q,p_2;P)\nonumber\\
&&\label{eqgenT3}\eea
In order to be as general as possible, we also give here formulas valid for any masses $\md$ and $\mup$.
In Eq.(\ref{eqgenT3}),  it is possible to perform the integration on the variable $s_q$ by deforming the
integration contour in the half plane $Re(s_q)>\gamma$. The only singularity in this domain is the pole $s_q=\varepsilon_{\uparrow}q$, and we end up, as usual, with an integral equation where $T_3^{\uparrow}$ in the integral term is evaluated "on the shell".
The next step is to make a change of variables and function according to
\bea
{\bf p}_1&=&{\bf p}'_1+\frac{\mup}{2\,\mup+\md}{\bf P}\\
{\bf p}_2&=&{\bf p}'_2+\frac{\mup}{2\,\mup+\md}{\bf P}\\
s'_P&=&s_P-\frac{{\bf P}^2}{2(2\,\mup+\md)}\\
t_3^{\uparrow}({\bf p}'_1,{\bf p}'_2;s'_P)&=& T_3^{\uparrow}(
\{{\bf p}_1,\varepsilon_{\uparrow}({\bf p}_1)\},
\{{\bf p}_2,\varepsilon_{\uparrow}({\bf p}_2)\}
;\{{\bf P},s_P\})\nonumber\\
\eea
This amounts to go to the center of mass reference frame and to evaluate $T_3^{\uparrow}$ "on the shell". ${\bf p}'_1$ is the momentum of a $\uparrow$ incoming atom in the center of mass reference frame (total impulsion ${\bf P}$ in "laboratory" frame).
$s'_P$ can be seen as the total "energy" minus the center of mass kinetic energy, {\it i.e}
the energy in the center of mass reference frame.
 In this way, we get
\begin{widetext}
\bea
t_3^{\uparrow}({\bf p}'_1,{\bf p}'_2;s'_P)
&=&
\frac{1}{\left[
\frac{(p'_1)^2 +(p'_2)^2}{2 m_r}
+\frac{{\bf p}'_1\cdot{\bf p}'_2}{m_{\downarrow}}
-s'_P
\right]}+
\int\frac{d^3{\bf q}'}{(2\pi)^3}
\frac{t_2(s'_P-\frac{(q')^2}{2 m_{A\uparrow,D}} )}{
\left[
\frac{(p'_1)^2 +(q')^2}{2 m_r}
+\frac{{\bf p}'_1\cdot{\bf q}'}{m_{\downarrow}}
-s'_P
\right]
}
t_3^{\uparrow}({\bf q}',{\bf p}'_2;s'_P)
\eea
\end{widetext}
We do not suppose here that the two species have equal masses.
 $m_r=\frac{m_{\uparrow}m_{\downarrow}}{(m_{\uparrow}+m_{\downarrow})}$ is the $2$-body reduced mass, and $m_{A\uparrow,D}=\frac{m_{\uparrow}(m_{\uparrow}+m_{\downarrow})}{(2 m_{\uparrow}+m_{\downarrow})}$ is the atom$\uparrow$-dimer reduced mass.
 
 This integral equation can be projected on the Legendre polynomials $P_l$. In this way, the integrals equation for different $l's$ decouple, and we find
 \begin{widetext}
 \bea
 t_{3,l}^{\uparrow}( p'_1,p'_2;s'_P)&=&-\frac{(2l+1)m_{\downarrow}}{p'_1 p'_2}
 Q_l\left(
 \frac{m_{\downarrow}(s'_P-
 \frac{(p'_1)^2 +(p'_2)^2}{2 m_r}
 )}{p'_1 p'_2}
\right)
 -\int_{0}^{+\infty}\frac{dq'}{2\pi^2} q'^2 
 t_2(s'_P-\frac{(q')^2}{2 m_{A\uparrow,D}})
 \frac{m_{\downarrow}}{p'_1 q'}
 Q_l\left(
 \frac{m_{\downarrow}(s'_P-
 \frac{(p'_1)^2 +(q')^2}{2 m_r}
 )}{p'_1 q'}
 \right)\nonumber\\
 &&\times t_{3,l}^{\uparrow}( q',p'_2;s'_P)\label{eqt3l}
 \eea
 \end{widetext}
 where $Q_l(z)=\frac{1}{2}\int_{-1}^{1}du\frac{1}{z-u}P_l(u)$ is the Legendre function of the second kind \cite{gradryz}.

 \section{Convolution integrals}\label{convol}
 We want to compute the following $4$-dimensional integral
 \begin{widetext}
 \bea
 I(\beta)&=&\int_{t_i>0}\Pi_{i=2}^{4} dt_i \,\theta\left(\beta-\left(t_2+t_3+t_4\right)\right)
 g\left(\beta-\left(t_2+t_3+t_4\right)\right)
 f_2(t_2)\,f_3(t_3)\,f_4(t_4)\label{eqI}
 \eea
 \end{widetext}
 We first integrate with respect to $t_2$, and use the convolution theorem for
 Laplace transforms
 \bea
 \int_0^t \,dt_2 g(t-t_2)\,f_2(t_2)&=& {\mathcal L}^{-1}\left(G(s_2)\,F_2(s_2)\right)
 \eea
 where $G(s)=\int_0^{+\infty}dt\,e^{s\,t}g(t)$ and $F_2(s)=\int_0^{+\infty}dt\,e^{s\,t}f_2(t)$. We use a notation such that the Laplace transforms of a given function is denoted by the corresponding capital letter. The symbol ${\mathcal L}$ denotes the Laplace transform, and ${\mathcal L}^{-1}$ the {\it inverse} Laplace transform.
 The integration with respect to $t_2$ therefore gives the contribution
 \bea
 g_2\left(\beta -\left(t_3+t_4\right)\right)&=&{\mathcal L}^{-1}\left(
 G(s)\,F_2(s)
 \right)_{t=\beta -\left(t_3+t_4\right)}
 \eea
 
 The integration on $t_3$  can be done similarly, and we get
 \bea
 g_3\left(\beta -t_4\right)&=&{\mathcal L}^{-1}\left(
 G_2(s)\,F_3(s)
 \right)_{t=\beta -t_4}\nonumber\\
 &=&{\mathcal L}^{-1}\left(
 G(s)\,F_2(s)\,F_3(s)
 \right)_{t=\beta -t_4}
 \eea
 
 The last integration on $t_4$ gives the full integral
 \bea
 I(\beta)&=&{\mathcal L}^{-1}\left(
 G_3(s)\,F_4(s)
 \right)_{t=\beta}\nonumber\\
 &=&{\mathcal L}^{-1}\left(
 G(s)\,F_2(s)\,F_3(s)\,F_4(s)
 \right)_{t=\beta}
 \eea
 In this way, we have transformed the $4$-dimensional integral Eq.\ref{eqI} in the $1$-dimensional following integral 
 \bea
 I(\beta)&=&\int_{{\mathcal C}_{\gamma}} \frac{d s}{2\pi\,i}e^{-\beta\,s}
 G(s)\,F_2(s)\,F_3(s)\,F_4(s)
 \eea
 with ${\mathcal C}_{\gamma}$ a contour parallel to the imaginary axis with $Re(s)=\gamma$, such that $G,F_2,F_3$ and $F_4$ are analytical for $Re(s)<\gamma$.
 
 This can be generalized easily to any $n$-dimensional integral of the form
 \begin{widetext}
  \bea
 I_n(\beta)&=&\int_{t_i>0}\Pi_{i=2}^{n} dt_i \,\theta\left(\beta-\left(t_2+\cdots+t_n\right)\right)
 g\left(\beta-\left(t_2+\cdots+t_n\right)\right)
  f_2(t_2)\,\cdots\,f_n(t_n)\label{eqI}
 \eea
 \end{widetext}
 Following the same method as before, we get
 \bea
 I_n(\beta)&=&\int_{{\mathcal C}_{\gamma}} \frac{d s}{2\pi\,i}e^{-\beta\,s}
 G(s)\,F_2(s)\,\cdots\,F_n(s)
 \eea

\end{document}